\documentclass[dvips,10pt,twocolumn,prd,superscriptaddress]{revtex4}

\usepackage{amsmath,amssymb,epsf,epsfig,amsthm,bm,graphicx,subfigure}

\newcommand{\tilq}{\widetilde{Q}}

\newcommand{\tilk}{\widetilde{K}}

\newcommand{\nwedge}{\!\wedge\!}
\newcommand{\bo}{{\omega_t}}
\newcommand{\bO}{{\Omega_t}}

\begin{document}

\title{`Mass without mass' from thin shells
in Gauss-Bonnet gravity }

\author{Elias Gravanis}
\affiliation{ Department of Physics, Kings College London}

\author{ Steven Willison}
\email{steve-at-cecs.cl}
 \affiliation{Centro de Estudios
Cient\'{\i}ficos (CECS) Casilla 1469 Valdivia, Chile }

\begin{abstract}
Five tensor equations are obtained for a thin shell in
Gauss-Bonnet gravity. There is the well known junction condition
for the singular part of the stress tensor intrinsic to the shell,
which we also prove to be well defined. There are also equations
relating the geometry of the shell (jump and average of the
extrinsic curvature as well as the intrinsic curvature) to the
non-singular components of the bulk stress tensor on the sides of
the thin shell.

The equations are applied to spherically symmetric thin shells in
vacuum. The shells are part of the vacuum, they carry no energy
tensor. We classify these solutions of `thin shells of
nothingness' in the pure Gauss-Bonnet theory. There are three
types of solutions, with one, zero or two asymptotic regions
respectively. The third kind of solution are wormholes. Although
vacuum solutions, they have the appearance of mass in the
asymptotic regions. It is striking that in this theory, exotic
matter is not needed in order for wormholes to exist- they can
exist even with no matter.

\end{abstract}

{\small Preprint: CECS-PHY-07/01}

\maketitle

\section{Introduction}

The dynamics of thin shells in Einstein's theory of gravity is
described by a set of five tensor equations~\cite{Israel:1966rt}.
One is an algebraic relation between the jump in the extrinsic
curvature and the intrinsic stress-energy tensor (the junction
condition). Four more relate the geometry of the shell (extrinsic
curvature on each side as well as the intrinsic curvature) to the
value of the bulk stress-energy tensor on the sides of the thin
shell. A fact about Einstein's theory is that if the intrinsic
stress-energy tensor of the shell vanishes then there is no jump
in the extrinsic curvature. This comes from the junction
condition.

In this paper, a similar analysis is performed for the Gauss-Bonnet
theory of gravity in five dimensions, which is quadratic in the
Riemann curvature. A qualitative difference is that the junction
condition~\cite{Davis-02, Gravanis-02} does not imply zero jump for
the extrinsic curvature when the energy tensor of the thin shell
vanishes~\cite{Meissner}. This is because the junction condition is
non-linear in the (extrinsic and intrinsic) shell curvatures. If the
bulk tensor also vanishes we obtain a non-smooth kink solution to
the vacuum field equations and can be thought of some kind of
soliton. Then the other four equations describe the dynamics of that
object. One of them implies the existence of a covariantly
conserved, symmetric tensor $\widetilde{Q}^a_{\ b}$ on the shell. It
involves the jump of the extrinsic curvature, whereas its
counterpart in Einstein theory doesn't.

These results are applied to a simple example. We consider
spherically symmetric shells in pure Gauss-Bonnet gravity. A
complete classification of non-null solutions with solitonic shells,
both static and time-dependent, is given. A particularly striking
type of solution is when two exterior regions are matched. Vacuum
thin shell wormhole solutions are found in which the stress tensor
on the shell is zero. The concept of `mass without
mass'~\cite{Misner-57} is shown to be realised in this context. The
exterior solution is that of the exterior of a massive object, but
the massive object is excised and replaced with another exterior
region connected by a wormhole throat which is a `thin shell of
nothingness'.

It is argued that these conclusions should also be true in
Einstein-Gauss-Bonnet and Lovelock gravity generally.
\\

\textbf{Notation:} Capital Roman letters $A$, $B$ etc. represent
five-dimensional tensor indices. Lower case Roman letters $a$, $b$
etc. represent four-dimensional tensor indices on the tangent space
of the worldsheet of the shell.

\subsection{Thin shells in Einstein's theory}

First, we review the formalism in General Relativity due to W.
Israel~\cite{Israel:1966rt}. Let $\Sigma$ be a hypersurface of
co-dimension 1 (the worldsheet of the shell) on which the stress
tensor is a delta function:
\begin{gather}\label{Stress}
 T_{AB} = \left(%
\begin{array}{cc}
  S_{ab} \delta(\Sigma) & \emptyset \\
  \emptyset & 0 \\
\end{array}%
\right)\, ,
\end{gather}
where $S^{a}_{\ b}$ is the intrinsic stress tensor on the shell. The
$\delta(\Sigma)$ is a Dirac delta function with support on the
shell. To simplify the presentation, we shall assume that tensors
are written in a basis $e_A = (e_a, n)$ which is adapted to the
shell so that $e_a$ are tangent vectors to $\Sigma$ and $n$ is
normal to $\Sigma$. The shell divides the space-time into two
regions, which are denoted by $M_+$ and $M_-$.

In Einstein's gravity, such a concentration of matter will produce a
discontinuity in the first derivative of the metric. This is given
covariantly by introducing the extrinsic curvature of the shell.
This is defined as $K_{ab} = e_a \cdot \nabla_{\!e_b} n$.
Intuitively, it measures the tangential rate of change of the normal
vector along the surface $\Sigma$.

The integration of the tangential-tangential components of the
Einstein equation across the infinitesimal width of the shell gives:
\begin{gather}\label{Israel1}
-\sigma(\Delta K^a_{b} - \delta^a_b\Delta K)
   = S^a_{b}\ ,
\end{gather}
where $\Delta K^a_{b}\equiv (K^+)^a_b - (K^-)^a_b$ is the jump in
the extrinsic curvature across the shell. The factor $\sigma = +1$
for a space-timelike shell (with a spacelike normal vector) and
$\sigma = -1$ for a space-like shell (timelike normal vector). The
projection of the normal-tangential components of the bulk Einstein
equation gives:
\begin{gather}
-\sigma(\Delta K^a_b - \delta ^a_b\Delta K)_{;a}
   = 0\ ,\label{Israel2}
\end{gather}
and\begin{gather} -\sigma(\tilk^a_b - \delta ^a_b\tilk)_{; a}
   = 0\ ,
\end{gather}
where $\tilk^{a}_{b} \equiv (K^+)^a_b + (K^-)^a_b$ is the sum of
the extrinsic curvature on each side of the shell. The semicolon
denotes the intrinsic covariant derivative on the shell. Also, the
projection of the normal-normal components of Einstein's equation
gives:
\begin{align}
\frac{\sigma}{4}
  \Delta K^a_c \tilk^b_d \,
  \delta^{cd}_{ab} & =0\ ,
\\
-R +\frac{\sigma}{4}\left\{
  \tilk^a_c \tilk^b_d + \Delta K^a_c
  \Delta K^b_d
\right\}\delta_{cd}^{ab} & = 0\ ,\label{Israel5}
\end{align}
where $R$ is the \emph{intrinsic} Ricci scalar of the shell. The
antisymmetrised Kronecker delta is defined as: $ \delta^{a b}_{cd} =
\delta^a_c \delta^b_d - \delta^b_c \delta^a_d$.

The first Israel junction condition (\ref{Israel1}) says that the
effect of the singular matter on the geometry is thus encoded in the
discontinuity of the extrinsic curvature. This condition can clearly
be inverted to determine the jump in the extrinsic curvature in
terms of intrinsic stress tensor. (Indeed, if one uses this to
replace $S^a_{b}$ in equations~\ref{Israel2}-\ref{Israel5}, the
expressions given in ref.~\cite{Israel:1966rt} are recovered). This
one-to one correspondence between $S^a_b$ and $\Delta K^a_b$ arises
because the Einstein equation is linear in the curvature.

\subsection{Thin shells in Gauss-Bonnet gravity}

A generalisation of Einstein's theory which is not linear in the
curvature was given by Lovelock~\cite{Lovelock:1971yv}. In five or
higher space-time dimensions it gives second order field equations.
Indeed, it is the most general second order metric theory of gravity
and can be thought of as natural correction to Einstein gravity in
more than four dimensions. In five dimensions the Lagrangian is:
\begin{gather}\label{EGB_action}
{\cal L} = c_0\, \sqrt{-g}\, d^5x + c_1\, {\cal R} \sqrt{-g}\, d^5x
+ c_2\, {\cal L}_{\text{GB}}.
\end{gather}
The first term is the cosmological constant, the second is the
Einstein-Hilbert term, the third is the so called Gauss-Bonnet term
which is quadratic in curvature:
\begin{gather*}
{\cal L}_\text{GB} : = \left({\cal R}^2 - 4 {\cal R}_{AB} {\cal
R}^{AB} + {\cal R}_{ABCD} {\cal R}^{ABCD} \right) \sqrt{-g}\, d^5x.
\end{gather*}
Above, the calligraphic script ${\cal R}$ is used to denote the five
dimensional curvatures.

Solutions with a hypersurface of codimension one with a Gauss-Bonnet
term have been studied extensively in the context of
brane-worlds~\cite{Kim,Meissner,Charmousis-02}, inspired by string
theory.
 A covariant junction condition, the analogue of
(\ref{Israel1}) was derived in Refs.~\cite{Davis-02,Gravanis-02}
using an action principle. Also, covariant equations of motion have
been derived from decomposition of the bulk field
equations~\cite{Maeda-03}. This approach provides an alternative
derivation of the junction condition. Also other covariant equations
for the brane have been derived in the literature, with the emphasis
being upon finding an effective theory of gravity on the brane. A
modified Einstein equation for the intrinsic metric on the
hypersurface has been obtained~\cite{Maeda-03}, in which there are
nonlinear corrections involving the extrinsic curvature and a
non-local piece coming from the Weyl tensor in the bulk. The
generalisation of (\ref{Israel2}), which says that the intrinsic
stress tensor on the hypersurface is covariantly conserved, is well
known~\cite{Davis-02}. Other works on equations of motion for branes
in Einstein-Gauss-Bonnet theory are Refs.~\cite{Barrabes-05}.

However, it seems that the analysis in the style of Israel's five
equations has not been done. Such a set of equations is of course
just an alternative formalism to that of Ref.~\cite{Maeda-03}, but
it shows some interesting information  which may be hidden in other
formalisms. In the next section, we present this analysis for a
shell embedded in a bulk in which the field equations for the pure
Gauss-Bonnet theory ($c_0=c_1=0$) hold. The analysis is given for
arbitrary matter on the shell and no matter in the bulk (the more
general case of Einstein-Gauss-Bonnet theory with matter in the bulk
is given in appendix \ref{GB}).
\\

Spherically symmetric shell solutions in vacuum in the five
dimensional Einstein-Gauss-Bonnet theory were considered in
Ref.~\cite{Charmousis-02}. These kind of solutions are of interest
in cosmology because there is a spatially homogeneous cosmological
metric induced on the shell, with an expansion factor governed by a
modified Friedmann equation. The solutions were restricted to
$\mathbb{Z}_2$ symmetry, where the metric on one side of the shell
is the mirror image of the other side. When this assumption is
dropped, solutions are generally very complicated~\cite{Konya}.
However, in Refs.~\cite{Crisostomo:2003qc,Dias}, general thin shells
in spherically symmetric spacetimes were examined for a certain
class of Lovelock theories. In those references a Hamiltonian
treatment of thin shells in GR~\cite{Olea} was generalised to
Lovelock gravity.

There is a curious possibility, which is not possible in
Einstein's theory. Because of the non-linearity in curvature, it
is possible to have a hypersurface where there is a discontinuity
in the extrinsic curvature without any stress tensor as source. In
other words, there is a thin shell made of nothing, where $S^a_b =
0$ but $\Delta K^a_b \neq 0$. These kind of solutions were
considered in Refs.~\cite{Meissner} (also an example in
11-dimensional Chern-Simons gravity was studied in Ref.
\cite{Hassaine}) and we shall call them `solitonic shells'. It
happens that, for these kinds of solutions, the junction
conditions can be resolved, without assuming $\mathbb{Z}_2$
symmetry, in a relatively simple way.

In section \ref{Matching_section}, explicit solitonic shell
solutions are found. We shall restrict ourselves to consider only
the Gauss-Bonnet term, i.e. the coefficients $c_0$ and $c_1$ shall
be set to zero. This theory, which we shall call pure Gauss-Bonnet
gravity, arises as the torsion-free sector of Chern-Simons theory of
the Poincare group in five dimensions
$ISO(4,1)$~\cite{Chamseddine:1989nu,Chamseddine:1990gk}. It can be
thought of as generalization of the interpretation of 2+1
dimensional General Relativity as a Chern-Simons theory for the
Poincare group~\cite{Achucarro:1987vz,Witten:1988hc}.

This theory has no Newtonian limit, so, like GR in three dimensions,
it should properly be regarded as a toy model for studying
qualitative features of gravity. The advantage for us is that the
spherically symmetric bulk solutions and the junction condition take
a very simple form. We are able to classify all of the spherically
symmetric solitonic thin shell solutions, without assuming
$\mathbb{Z}_2$ symmetry.

 Here, the focus shall not be on cosmology on the shell. The
main interest will be in wormhole solutions which behave in a
sense like material particles even though they are not massive
solutions. That is, instead of being the universe, the shell
should perhaps be thought of as a kind of particle.

Although only the pure Gauss-Bonnet theory is considered explicitly,
we comment on the generalization to general Lovelock theory in
section \ref{Comments}.

\section{The five Equations for a shell in Gauss-Bonnet gravity}
\label{Five_Section}

Let us for now concentrate on the Gauss-Bonnet term, setting $c_0$
and $c_1$ to zero and $c_2=1$ in the action (\ref{EGB_action}). The
field equation of pure Gauss-Bonnet gravity is:
\begin{gather}\label{GB_field_equation_tensors}
 - \frac{1}{8}
  \delta^{A C_1 \cdots C_4}_{ B D_1 \cdots D_4}
 \mathcal{R}^{D_1 D_2}_{\ \ C_1 C_2} \mathcal{R}^{D_3 D_4}_{\ \ C_3 C_4}
 = T^A_{\ B}\ .
\end{gather}

Let us find the analogue of Israel's five equations
(\ref{Israel1}-{\ref{Israel5}) for Gauss-Bonnet gravity. Since the
origin of these equations is clear, we shall just state here the
results. The proof is given in Appendix \ref{GB}.

Here the results are summarised for the case where the bulk energy
tensor is zero. First we define the following symmetric tensor:
\begin{gather}\label{def_Q_tensor}
 Q^a_{\ b} \equiv K^c_f \left(2\sigma R^{de}_{gh}
 -\frac{4}{3} K^d_g  K^e_h \right)
 \delta_{b\, c\, d\, e}^{afg\, h}.
\end{gather}
Also we define $\Delta Q^a_b \equiv (Q^+)^a_b -(Q^-)^a_b$, the jump
across the shell and $\tilq^a_b \equiv (Q^+)^a_b +(Q^-)^a_b$, the
sum of $Q^a_b$ evaluated on each side.

The integration of the tangential-tangential components of the field
equation (\ref{GB_field_equation_tensors}) across the infinitesimal
width of the shell gives the junction
condition~\cite{Davis-02,Gravanis-02}:
\begin{gather}\label{GB_Junction}
\Delta Q^a_{\ b} = -2 S^a_b \, ,
\end{gather}
The projection of the normal-tangent components of the bulk field
equations onto the shell tell us that the intrinsic stress tensor is
covariantly conserved~\cite{Davis-02}:
\begin{gather}\label{conservation_dif_Q}
\Delta Q^a_{\ b; a} = 0 \quad \Rightarrow \quad S^a_{\ b; a} = 0\ ,
\end{gather}
and also that the tensor $\tilq^a_{\ b}$ is covariantly conserved on
the shell:
\begin{gather}
\tilq^a_{\ b;a} = 0\ .\label{conservation_av_Q}
\end{gather}

The projection of the normal-normal component of the field equation
gives:
\begin{widetext}
\begin{align}
-\frac{3}{8}\left\{ \tilk^a_b \Delta Q^b_{\ a} + \Delta K^a_b
\tilq^b_{\ a}\right\} +\frac{\sigma}{2} \tilk^a_e \Delta K^b_f
R^{cd}_{\ \ gh}\, \delta_{a\, b\, c\, d}^{e\,fg\, h} & = 0\,
,\label{Long_eqn}
\\
\frac{1}{2} R^{ab}_{\ \ ef} R^{cd}_{\ \ gh}\, \delta^{efgh}_{abcd}
-\frac{3}{8}\left\{ \tilk^a_b \tilq^b_{\ a} + \Delta K^a_b \Delta
Q^b_{\ a}\right\} +\frac{\sigma}{4} \left\{ \tilk^a_e \tilk^b_f +
\Delta K^a_e \Delta K^b_f\right\} R^{cd}_{\ \ gh} \delta_{a\, b\,
c\, d}^{e\,fg\, h}  & = 0 \,.\label{Very_Long_eqn}
\end{align}
\end{widetext}

Equations (\ref{GB_Junction}-\ref{Very_Long_eqn}) are the five
equations characterising the shell. The first two are already known.
The last two are rather complicated and perhaps not very useful in
describing shells (although they may be useful in the Hamiltonian
formalism of Gauss-Bonnet gravity - see below). On the other hand,
equation (\ref{conservation_av_Q}) has some surprising consequences
which have gone unnoticed.

Because of the non-linearity of the Gauss-Bonnet theory, one can not
solve algebraically for the jump in extrinsic curvature in terms of
the intrinsic stress tensor. There are two independent quantities
$\Delta Q^a_{\ b}$ and $\tilq^a_{\ b}$, which can be expressed as:
\begin{gather*}
 \Delta Q^a_{\ b} = \Delta K^c_f \left(2\sigma R^{de}_{gh}
 -\frac{1}{3}\Delta K^d_g \Delta K^e_h - \tilk^d_g \tilk^e_h \right)
 \delta_{b\, c\, d\, e}^{afg\, h},
 \\
 \tilq^a_{\ b} = \tilk^c_f \left(2\sigma R^{de}_{gh}
 -\frac{1}{3}\tilk^d_g \tilk^e_h - \Delta K^d_g \Delta K^e_h \right)
 \delta_{b\, c\, d\, e}^{afg\, h}.
\end{gather*}
These quantities both depend nonlinearly  on $\Delta K^a_{\ b}$.
Only one of these is determined by the stress tensor, \emph{but both
are covariantly conserved.}

Note that when the surface $\Sigma$ is spacelike, $Q^a_b$ arises
naturally in the Hamiltonian formalism. It is proportional to the
momentum canonically conjugate to the spatial metric. Equations
(\ref{conservation_dif_Q}) and (\ref{conservation_av_Q}) say that
the extrinsic curvature can jump in a way that conserves the
constraint ${\cal H}_a = 0$. Equations (\ref{Long_eqn}) and
(\ref{Very_Long_eqn}) say that any discontinuity must preserve the
constraint ${\cal H}_{\perp} = 0$. Expressions for ${\cal H}_a$ and
${\cal H}_\perp$ in Lovelock gravity were first given in Ref.
\cite{Teitelboim}. The dynamical part of the field equations in
vacuum says that a discontinuity must obey $\Delta Q^a_b =0$.

The above five equations are for the pure Gauss-Bonnet theory. For
the more general action (\ref{EGB_action}), the generalisation is
straightforward. It is simply a linear combination of the terms
appearing in the Israel equations with the those of the
Gauss-Bonnet. This will be given explicitly in the appendix in eqns
(\ref{EGB_first}- \ref{EGB_fifth}).

\section{Solitonic spherical shells in pure Gauss-Bonnet gravity}
\label{Matching_section} Let us consider the pure Gauss-Bonnet
theory, with just the quadratic Lovelock term in the action. This
choice remains largely unstudied, no doubt because it does not
include the Einstein Hilbert term. The theory is not in any sense a
small correction to General Relativity. However, it offers a useful
toy model in which to study thin shells, finding exact solutions. In
this section, solitonic shells are found in spherically symmetric
background.

It is useful to use differential form notation. A brief explanation
of this is given in Appendix \ref{Defs}. In this notation, the field
equation is:
\begin{gather}\label{GB_field_equation}
c_2 \Omega^{AB}\nwedge \Omega^{CD}\epsilon_{ABCDF} = -2T_F\, ,
\end{gather}
where $\Omega^{AB}$ is the curvature two-form and $T_A$ is the
stress-energy four-form.

The spherically symmetric vacuum solution is:
\begin{gather}\label{spherically_symmetric}
ds^2 = - dt^2 + \frac{dr^2}{\beta^2} +r^2 d\Omega^2.
\end{gather}
Here $d\Omega^2$ is the line element of the unit three-sphere. This
is a special case of the solution of Boulware and
Deser~\cite{Boulware} for Einstein-Gauss-Bonnet, $\beta$ being the
constant of integration. This space-time was discussed recently in
Ref.~\cite{Cai:2006pq} but we are not aware of any previous detailed
study of this metric in the literature.

A basis is chosen such that the vielbein and spin connection take
the form:
\begin{gather}\label{vielbeins}
E^0 = dt, \quad E^1 = dr/\beta, \quad E^{i} = r \widetilde{E}^i,
\\\label{connections}
\omega^{1}_{\ i} = -\beta\widetilde{E}^i, \quad \omega^i_{\ j} =
\widetilde{\omega}^i_{\ j}\ .
\end{gather}
{\bf Notation:} $\widetilde{E}^i$ and $\widetilde{\omega}^i_{\ j}$
are the intrinsic vielbeins and spin connection on the three sphere.
The lower case Latin indices from the middle of the alphabet run
from $2$ to $4$.

Note that, although the constant factor $\beta$ looks innocuous,
\emph{this space-time is not locally flat}. The non-vanishing part
of the curvature two-form is:
\begin{gather}\label{curvature}
\Omega^{ij} = (1 - \beta^2) \widetilde{E}^i \nwedge \widetilde{E}^j.
\end{gather}
The five-dimensional Ricci scalar is
${ \cal R} = 6(1-\beta^2)/r^2$. For $\beta \neq 1$ there is clearly
a curvature singularity at $r = 0$.

Even though the curvature diverges, the singularity is
well-behaved in the sense that the field equations are
well-defined under integration. From (\ref{GB_field_equation}) the
energy tensor is zero everywhere outside the origin $r=0$. The
mass can be calculated unambiguously by integrating the field
equations in a ball centred around $r = 0$, using the Gauss-Bonnet
theorem. The result is:
\begin{gather}\label{mass}
m =16 \pi^2 c_2 \left[ 1 - \frac{|\beta|}{2}(3 -\beta^2) \right]\ .
\end{gather}
%This agrees with the mass calculated using Noether's theorem (see
%e.g. Ref. \cite{Mora}).
\begin{figure}[t]
  \begin{center}
    {\includegraphics[width =0.45\textwidth]{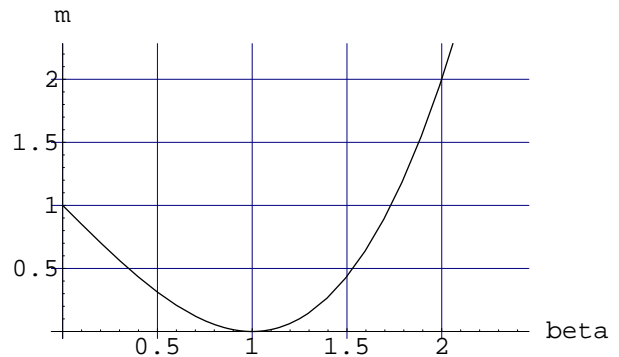}}
    \caption{\small{The mass as a function of $|\beta|$ in units such
    that $16\pi^2 c_2=1$}. Note that $m(\sqrt{3})=m(0)$ and
$m(2)=2$. The mass is non-negative for
    all values of $|\beta|$. For $0 < m \le 1$ there are two values of
$|\beta|$ producing the same mass.}
    \label{Mass_vs_beta}
  \end{center}
\end{figure}
It is natural to take $c_2$ to be positive, so that all solutions
have positive mass (see Fig. \ref{Mass_vs_beta}). For convenience, a
choice of units is made so that $16\pi^2 c_2=1$.

\subsection{Non-smooth static vacuum solutions}
\label{static}

Let us try to match two different point particle metrics on a
timelike hypersurface $\Sigma$ at a constant radius. The surface
divides space-time into two regions, $M_+$ and $M_-$. The metric in
each region is given by:
\begin{gather}\label{plus_metric}
ds^2_+ = -dt^2 + \frac{dr^2_+}{\beta_+^2} + r_+^2 d\Omega^2,
\\\label{minus_metric}
ds^2_- = -dt^2 + \frac{dr^2_-}{\beta_-^2} + r_-^2 d\Omega^2.
\end{gather}
and the hypersurface is located at $r_+ = r_- = r_0$ a constant so
that the induced metric is continuous.

For a static shell, the choice of vielbeins (\ref{vielbeins})
provides a frame adapted to $\Sigma$, i.e. $E^a = (E^0, E^i)$ are
dual to the intrinsic frame on $\Sigma$ and $E^1$ is dual to the
normal vector, $n = \beta_\pm \partial_{r_\pm}$. The normal vector
is, by convention, chosen to point from $M_+$ to $M_-$. Note that,
in our conventions, the orientation of the embedding of $\Sigma$
into $M_\pm$ is determined by the sign of $\beta_\pm$. There are
three choices:
\\
\textbf{Type I:} If $\beta_-$ and $\beta_+$ are both positive, the
global structure is the same as for the smooth solution. The radial
coordinate in $M_-$ is decreasing as one moves away from $\Sigma$
and the radial co-ordinate in $M_+$ increases as one moves away from
$\Sigma$. The region $M_-$ is the interior and contains the point
singularity. The region $M_+$ is the exterior. If $\beta_+$ and
$\beta_-$ are both negative, the global structure of the spacetime
is the same but with the roles of $M_+$ and $M_-$ swapped.
\\
\textbf{Type II$\!$ a):} If $\beta_-$ is positive and $\beta_+$
negative, two interior regions are joined together to form a
spatially closed universe. Each region contains a point source.
\\
\textbf{Type II$\!$ b):} If $\beta_-$ is negative and $\beta_+$ is
positive, two exterior regions are joined together. There are two
asymptotic regions $r_+ \to \infty$ and $r_- \to \infty$ and no
point sources. These are wormhole space-times.
\begin{figure}[ht]
     \centering
     \subfigure {
          \includegraphics[width=.35\textwidth]{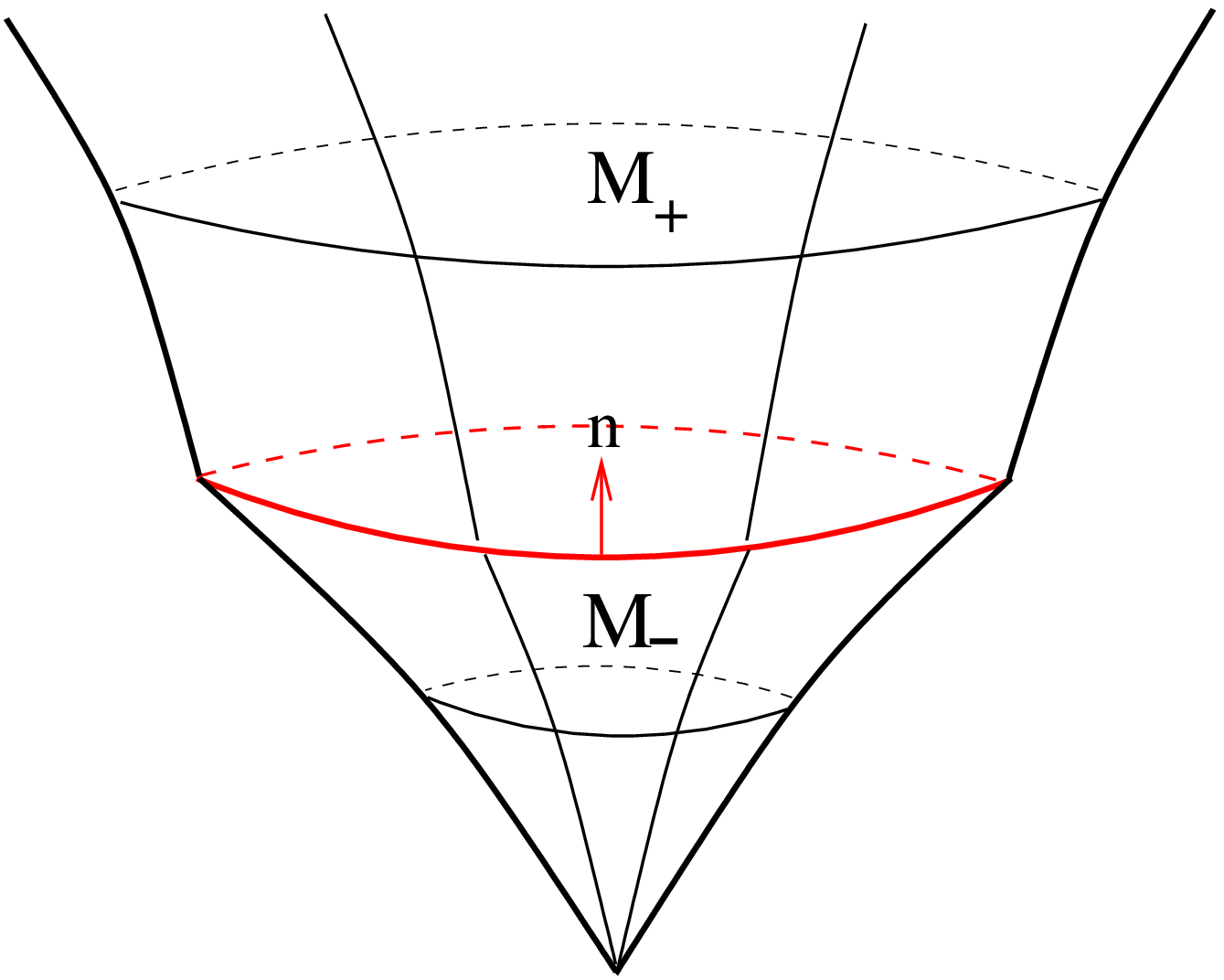}}\\
          {\bf Type I}. sign($\beta_+) =$ sign $(\beta_-)$.
     \\
     \vspace{.3in}
     \subfigure{
           \includegraphics[width=.38\textwidth]{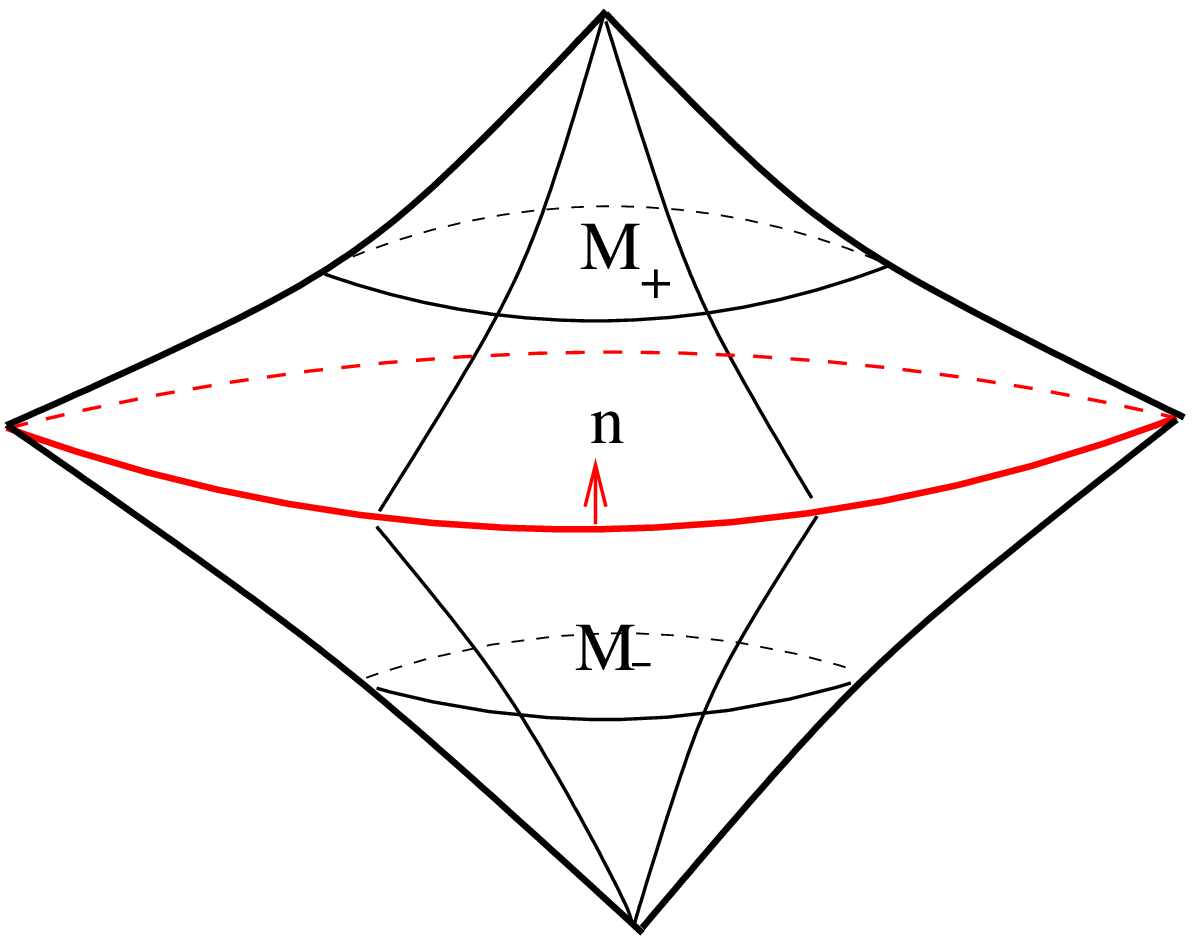}}\\
           {\bf Type IIa)}. $\beta_-$ positive and $\beta_+$ negative.
     \\
     \vspace{.3in}
     \subfigure {
           \includegraphics[width=.38\textwidth]{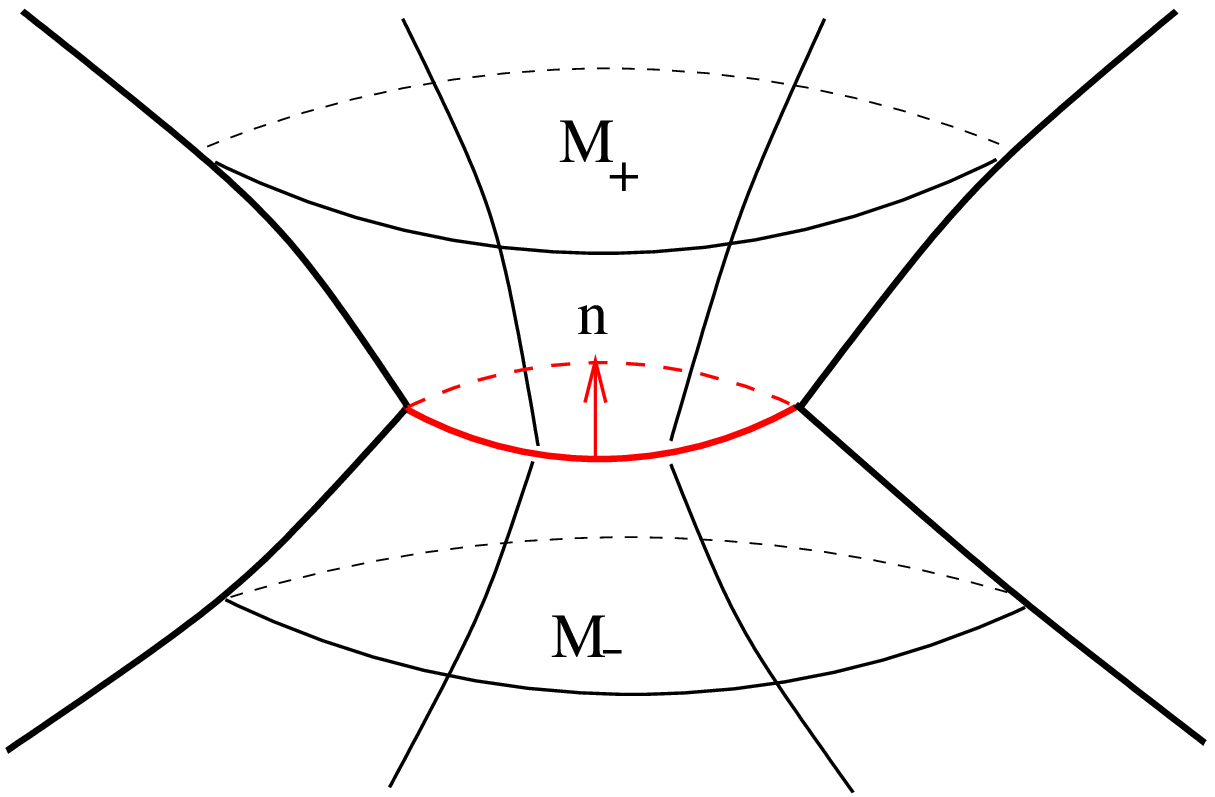}}\\
           {\bf Type IIb)}. $\beta_-$ negative and $\beta_+$
           positive.
      \caption{The three types of matching. The time direction is suppressed.
 Also two spatial dimension are suppressed so that 3-spheres are represented
 by circles.}
\end{figure}

If $\beta_+ \neq \beta_-$ the spin connection is discontinuous,
corresponding to a singular curvature. In Einstein's theory, such a
discontinuity on a time-like surface could never be a vacuum
solution. However, in the Gauss-Bonnet theory, the matching
conditions (\ref{GB_Junction}) tell us that the stress-energy tensor
located on the hypersurface vanishes if
\begin{gather*}
Q^+_a - Q^-_a  =0 \ ,
\end{gather*}
where $Q_a$ is defined in equation (\ref{def_Q}).

The only non-zero components of the second fundamental form
  are $\theta^{1i}
=-\beta \widetilde{E}^{i}$. The intrinsic geometry of the
hypersurface is $\mathbb{R}^1 \times S^3$ so the non-vanishing
components of intrinsic curvature is $\Omega_{\|}^{ij} =
\widetilde{E}^i \wedge\widetilde{E}^j$. The only component of $Q_a$
is:
\begin{gather*}
Q_ 0 = - 4 \beta \left( 1- \frac{1}{3}\beta^2 \right)
\widetilde{E}^i \wedge \widetilde{E}^j \wedge \widetilde{E}^k
\epsilon_{01ijk}\ .
\end{gather*}
[In tensor language, the only components of the extrinsic curvature
are $K^i_j =- \beta \delta^i_j$ and the only component of $Q^a_b$ is
$Q^0_0 = - 4\, 3! \beta \left( 1- \frac{1}{3}\beta^2 \right)$.]

The junction condition reduces to:
\begin{gather}
3\beta_+ -
\beta^3_+ = 3\beta_- -\beta_-^3\ .\label{condition_betas}
\end{gather}
There are two useful alternative ways to use equation
(\ref{condition_betas}). We can use it either to derive a junction
condition in terms of the metric parameters $\beta_\pm$ or a
condition in terms of the masses. Let us first find the condition in
terms of $\beta_\pm$. We note that (\ref{condition_betas})
factorises to give either $\beta_+ - \beta_- =0$ (which is trivial-
the metric is matched smoothly) or
\begin{gather}
\beta_+^2 + \beta_-^2 + \beta_+ \beta_- - 3 = 0.
\end{gather}
The above tells us which metrics can be matched together at a static
surface of constant $r$. It clearly has non-trivial solutions
% if
%we parametrize $\beta_+=R \cos\psi$ and $\beta_-=R \sin \psi$ we
%have
%\begin{equation}\label{}
%\beta_+=\frac{\sqrt{6}\cos\psi}{\sqrt{2+\sin 2\psi}} \quad , \quad
%\beta_-=\frac{\sqrt{6}\sin\psi}{\sqrt{2+\sin 2\psi}} \ .
%\end{equation}
%\textbf{Alternatively, we could parameterise the solutions by:}
%\begin{equation}
% \beta_+ = \sqrt{3} \sin \varphi + \cos \varphi,\qquad
% \beta_+ = -\sqrt{3} \sin \varphi + \cos \varphi
%\end{equation}
which are described by an ellipse in the parameter space $(\beta_+,
\beta_-)$.
\begin{figure}[ht]
  \begin{center}
    {\includegraphics[width=.4\textwidth]{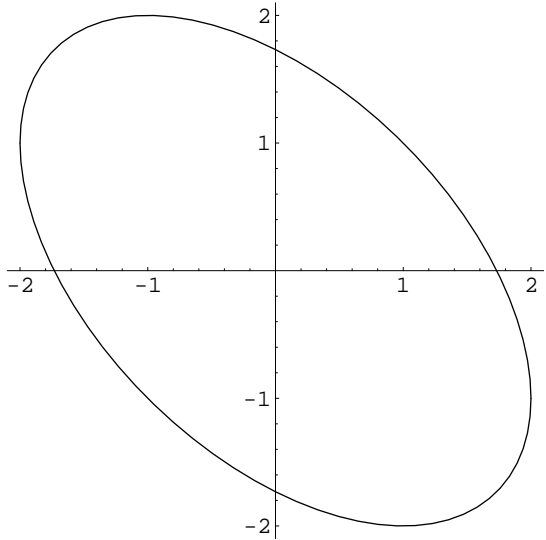}}
    \caption{\small{The static vacuum solutions of the junction
    condition describe an ellipse in the space of $\beta_+$ and
    $\beta_-$.
    The top right and bottom left quadrants
    correspond to a matching of type I with standard orientation.
    The top left quadrant corresponds to solutions of type IIa,
     the ``closed universe".
    The bottom  right quadrant corresponds to solutions of type
  IIb, wormholes.  }}
   \label{ellipse}
  \end{center}
\end{figure}

Alternatively, we can use formula (\ref{mass}) to express
(\ref{condition_betas}) in terms of the masses.
There are two cases: \\
Type I) The masses must be the same:
\begin{gather}\label{masses_same}
m_+ = m_- \quad \text{if} \ \text{sign}(\beta_+) =
\text{sign}(\beta_-).
\end{gather}
\\
Type II) There is a condition on the sum of the masses:
\begin{gather}\label{sum_of_masses}
m_+ + m_- = 2 \quad \text{if} \quad \text{sign}(\beta_+) =
-\text{sign}(\beta_-).
\end{gather}

As shown in figure \ref{Mass_vs_beta}, the cubic form of the mass
formula (\ref{mass}) leads to an ambiguity in the metric. If the
mass is in the range $ 0<m< 1$, there are two possible solutions-
one with $\beta$ between 0 and 1 and another solution with $\beta $
between 1 and $\sqrt{3}$. Thus, there exist non-smooth matchings of
type I) for masses less than $1$.

The matchings of type II) can occur if one of the $\beta$'s is
between 0 and $\sqrt{3}$ and the other is between $\sqrt{3}$ and
$2$, i.e. for masses less than $2$.

\subsection{Non-smooth time-dependent solutions}
\label{non-static}

In the previous example, we matched vacuum solutions at a static
surface. It is also possible to match at a surface which is not
static. We shall restrict ourselves to solutions which respect the
spherical symmetry of the smooth solutions: the hypersurface is
defined by $r$ some function of $t$. There are three possibilities:
timelike, spacelike or null surfaces. Here only the timelike and
spacelike cases will be considered.

\subsubsection{Timelike} There are two regions with metrics given by
(\ref{plus_metric}) and (\ref{minus_metric}). They are joined at a
hypersurface which is an expanding or shrinking 3-sphere.  The
hypersurface $\Sigma$ is defined parametrically by $r_- = a(\tau)$,
$r_+ = a(\tau)$ and $t_+ = T_+(\tau)$, $t_- = T_+(\tau)$. Here $a$
is some function of $\tau$, a time coordinate on $\Sigma$. The
induced metric is:
\begin{gather*}
ds^2_{\Sigma_\pm} = -d\tau^2 \left(\dot{T}_\pm^2 -
\frac{\dot{a}^2}{\beta_\pm^2}\right) +a^2(\tau) d\Omega^2.
\end{gather*}
We require that $ds^2_{\Sigma_+} = ds^2_{\Sigma_-}$ for a continuous
metric. Note that $a$ must be the same function on both sides
because it is the radius of curvature of the 3-sphere. It is natural
to choose $\tau$ to be the proper time co-ordinate on $\Sigma$ so
that $\dot{T}_\pm^2 - \frac{\dot{a}^2}{\beta_\pm^2} = 1$ and
\begin{gather*}
ds^2_\Sigma = -d\tau^2 +a^2(\tau) d\Omega^2\ .
\end{gather*}

In order to evaluate the junction conditions, we introduce a frame
adapted to $\Sigma$. The intrinsic vielbeins are $(E^{\hat{0}},
E^i)$, where $E^{\hat{0}}= d\tau = $ and the normal is
$E^{\hat{1}}$. They are related to the vielbeins (\ref{vielbeins})
by the Lorentz transformation:
\begin{gather*}
\left(%
\begin{array}{c}
  E^{\hat{0}} \\\\
  E^{\hat{1}} \\
\end{array}%
\right) = \left(%
\begin{array}{cc}
  \sqrt{1+\dot{a}^2/\beta_\pm^2} & -\dot{a}/\beta_\pm \\
  -\dot{a}/\beta_\pm & \sqrt{1+\dot{a}^2/\beta_\pm^2} \\
\end{array}%
\right)
\left(%
\begin{array}{c}
  E^{0}_\pm \\\\
  E^{1}_\pm \\
\end{array}%
\right) .
\end{gather*}
The vielbeins tangent to the 3-sphere, $\widetilde{E}^i$, are not
transformed. In this basis, the second fundamental form is:
\begin{gather*}
\theta_\pm^{\hat{1}i} = - \sqrt{\beta_{\pm}^2+\dot{a}^2}\
\widetilde{E}^i, \qquad \theta_\pm^{\hat{1}\hat{0}} =
-\frac{\ddot{a}}{\sqrt{\beta_{\pm}^2+\dot{a}^2}}\ d\tau\ .
\end{gather*}
There are now two nonzero equations coming from the junction
conditions, but they are not independent. The first is $\Delta
Q_{\hat{0}} =0$ which gives:
\begin{gather}\label{solution non squared}
\left[\text{sign}(\beta)\sqrt{\beta^2+\dot{a}^2}\left(
1+\frac{2}{3}\dot{a}^2 - \frac{1}{3}\beta^2 \right) \right]^+_- =
0\, ,
\end{gather}
where the square bracket $[\cdots]^+_-$ denotes the difference in
the argument evaluated on each side of $\Sigma$. The other equation,
$\Delta Q_i =0$ gives simply the derivative with respect to $\tau$
of the first.

Squaring (\ref{solution non squared}) and solving gives:
\begin{equation}\label{general rho dot}
\dot a^2=\frac{(\beta_+^2+\beta_-^2 +\beta_+ \beta_-
-3)(\beta_+^2+\beta_-^2 -\beta_+ \beta_-
-3)}{3(\beta_+^2+\beta_-^2-2)} \ .
\end{equation}
[It can be checked separately using (\ref{solution non squared})
that $\beta_+^2 + \beta_-^2=2$ is not a solution except in the
trivial case where $\beta_+ = \beta_- =1$. To obtain (\ref{general
rho dot}) we have divided through by a common factor of $(\beta_+^2-
\beta_-^2)$. In principle, the case $\beta_+^2 - \beta_-^2 = 0 $
should also be verified separately. However, it turns out that this
case is correctly described by equation (\ref{general rho dot}) and
the set of inequalities given below.] Time-dependent solutions only
exist when $\dot{a}^2 > 0$. This occurs when an odd number of the
three inequalities:
\begin{gather}
 \beta_+^2+\beta_-^2 +\beta_+ \beta_-
-3 > 0 \, ,\nonumber
\\
 \beta_+^2+\beta_-^2 -\beta_+ \beta_-
-3 > 0\, ,\label{Consistency1}
\\\nonumber
 \beta_+^2+\beta_-^2-2 > 0\, ,
\end{gather}
are satisfied. Since we have squared the junction condition, we must
plug the solution back into (\ref{solution non squared}) in order to
determine the relative orientation consistent with the solution.
%This consistency can  be expressed as:
%\begin{gather}
% \text{sign} (\beta_+ \beta_-) =
% \text{sign} \left[\left(
% 1+\frac{2}{3}\dot{a}^2 - \frac{1}{3}\beta_+^2 \right) \left(
% 1+\frac{2}{3}\dot{a}^2 - \frac{1}{3}\beta_-^2 \right)\right]
% \\= -\text{sign} \frac{(\beta_+ + \beta_-)^2(\beta_+ - \beta_-)^2
% (2\beta_+^2 + \beta_-^2 -3)(\beta_+^2 + 2\beta_-^2
% -3)}{81(\beta_+^2 + \beta_-^2 -2)^2}\, .
%\end{gather}
A consistent solution will obey an even number of the three
inequalities:
\begin{gather}
 \beta_+\beta_- > 0 \, ,\nonumber
\\
 2\beta_+^2+\beta_-^2 - 3 > 0\, ,\label{Consistency2}
\\
 \beta_+^2+ 2\beta_-^2- 3  > 0\, .\nonumber
\end{gather}

\begin{figure}[t]
  \begin{center}
    {\includegraphics[width=.49\textwidth]{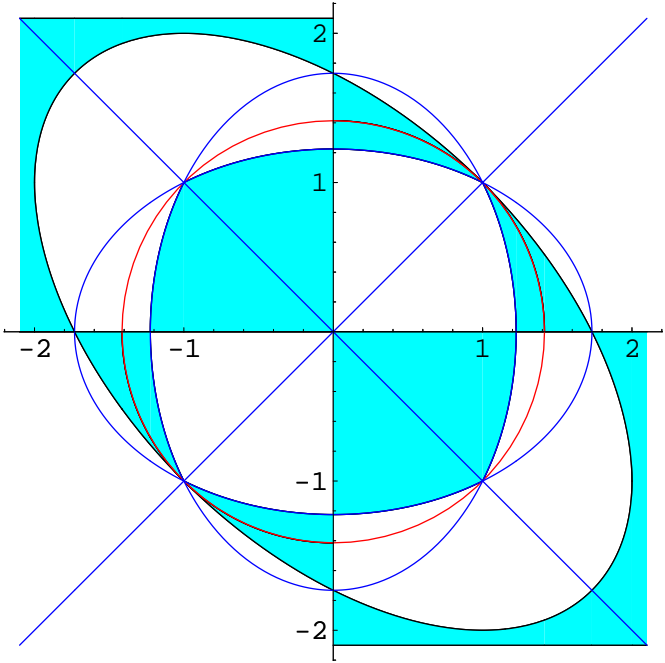}}
    \caption{\small{The shaded regions are the allowed solutions.
    The circle of radius $\sqrt{2}$ divides the solutions where
    $\Sigma$ is spacelike (inside) from those where $\Sigma$ is
    timelike (outside). The solutions where $\Sigma$ is static lie
    on the large ellipse. The top left and bottom right quadrants are
    shaded everywhere outside this ellipse.}}
    \label{magic_ellipses}
  \end{center}
\end{figure}

\subsubsection{Spacelike}

Following a similar analysis for the case of a spacelike surface
gives the junction condition:
\begin{gather}\label{solution non squared_spacelike}
\left[\text{sign}(\beta)\sqrt{\dot{a}^2-\beta^2} \left(
\frac{2}{3}\dot{a}^2 - 1 + \frac{1}{3}\beta^2 \right) \right]^+_- =
0\, .
\end{gather}
This can be squared and solved for $\dot{a}$ to give:
\begin{gather}\label{solution_squared_spacelike}
\dot a^2 = -\frac{(\beta_+^2+\beta_-^2 +\beta_+ \beta_-
-3)(\beta_+^2+\beta_-^2 -\beta_+ \beta_-
-3)}{3(\beta_+^2+\beta_-^2-2)} \ .
\end{gather}
Inserting this value of $\dot{a}^2$ back into (\ref{solution non
squared_spacelike}), the consistency of the solution tells us that
an even number of the inequalities (\ref{Consistency2}) must be
satisfied. Also, the condition that the square root be real gives:
\begin{gather}
\beta_+^2+\beta_-^2-2 < 0\, .
\end{gather}
The set of inequalities for spacelike and timelike shells is
depicted graphically in Fig.~\ref{magic_ellipses}}.
\\

The solutions with $\Sigma$ spacelike represent a breakdown of
determinism. The extrinsic curvature can jump instantaneously from
one value to another in a way which is not predicted by the initial
conditions. Note that, restricting ourselves to spherically
symmetric metrics, such jumps are ruled out for masses greater than
$1$ (i.e. $\beta > \sqrt{2}$).

\subsection{`Mass without mass' and conserved quantities}

Let us consider the solutions of type IIb), the wormholes. These
solutions contain no point sources, i.e. the stress tensor is
everywhere zero. An observer in the region $M_+$ feels a spacetime
as if there were a spherically symmetric source of mass $m_+$ on
the other side of the shell. If he moves across the shell, instead
of accessing a source he feels a mass $m_-$ behind him. These
wormholes illustrate the concept of `mass without mass'. The
non-trivial topology of the vacuum solution creates the illusion
of having a massive particle.

In section \ref{Five_Section} it was shown that there are two
covariantly conserved symmetric tensors on the shell, $\Delta Q^a_b$
and $\tilq^a_b$. Equivalently, this can be stated that $(Q^+)^a_b$
and $(Q^-)^a_b$ are covariantly conserved independently of each
other. In certain cases then, such as if there exists a Killing
vector on the hypersurface, we can define a conserved quantity
associated with $\tilq^a_b$.

Recall the static and non-static solutions of section
\ref{non-static}. In the non-static solutions the vector $e_{\hat 0}
=
\partial_\tau$ is not a Killing vector on the hypersurface. However,
it is still true that $ i^*d Q^{\pm}_{\hat 0} = 0$, so the
quantities $ \int_{S^3} Q^{\pm}_{\hat 0}$ are conserved with $\tau$,
where $S^3$ is any 3-sphere given by $\tau$=constant. The vacuum
matching amounts to $Q^+_a-Q^-_a=0$.
%There is only one non-trivial
%covariantly conserved quantity,
Note that $\tilq_a \equiv Q^+_a + Q^-_a =2 Q^+_a=2 Q^-_a$. We define
the conserved quantity:
\begin{equation}\label{conserved}
 \widetilde{q} \equiv \frac{\, c_2}{2}\int_{S^3} \tilq_{\hat 0}\, .
\end{equation}

 In the static case one finds for the wormhole solutions
\begin{equation}\label{explicit static conserved}
 \widetilde{q} =  m_+ - m_- = 2m_+ -2\, .
\end{equation}
This is what the total energy of the shell would have been, as
measured from $M_+$, had the two regions been matched with the other
orientation, i.e. with $M_-$ replaced by an interior region of mass
$m_-$. The sign of this charge is somewhat arbitrary. An observer in
$M_+$ would naturally define it with a plus sign but an observer in
$M_-$ with a minus sign.

%As we see clearly from (\ref{explicit static conserved}) the
%conserved quantity is a measure of asymmetry between the $-$ and $+$
%side across the shell.

Consider now the non-static wormhole solutions with a timelike shell
$\Sigma$ described in subsection \ref{non-static}. We have
\begin{equation*}\label{}
(Q^+)^{\hat{0}}_{\hat 0}=-4\ 3! \ \sqrt{\beta_+^2 + \dot a^2}
\Big(1+\dot a^2-\frac{1}{3}\big(\beta_+^2 + \dot a^2\big) \Big).
\end{equation*}
The other components vanish (by $\dot a=$constant). A similar
formula holds for $Q_{\hat 0}^-$. For the general case the result is
rather untransparent. Consider the case where the metric in $M_-$ is
flat i.e. $\beta_-=-1$. The result simplifies considerably and one
finds, using (\ref{general rho dot}),
\begin{equation*}\label{}
 \widetilde{q} =  2\
\Big(\frac{\beta^2_+-1}{3}\Big)^{3/2}.
\end{equation*}
Expressed in terms of the speed $v = dr/dt_-$ measured by the
Minkowski observer, this reads:
\begin{equation*}\label{}
 \widetilde{q} = 2\ (1-v^2)^{-3/2}.
\end{equation*}
Note that in the static case $v=0$ we have $m_+=2$ from the formula
(\ref{sum_of_masses}). The non-static result is modified by the
inverse relativistic factor of the volume of the isotropically
expanding 3-sphere.
\\

 As noted above $\widetilde{q}$ tells us about the asymmetry
of the vacuum wormhole. It is interesting that in the static case
it vanishes for unit masses $m_+=m_-=1$. One can check that
$\widetilde{q}$ vanishes in the non-static wormhole too when
$m_+=m_-$. More generally, in the nonstatic case, $\widetilde{q}$
vanishes on the small ellipses (blue lines) shown in figure
\ref{magic_ellipses}. Also $\widetilde{q}$ goes to infinity at the
(red) circle $\beta_+^2 + \beta_-^2 = 2$ (and also at $\beta_+ \to
\infty$ and $\beta_- \to \infty$). It is tempting to conjecture
that $\widetilde{q}$ is a kind of gravitational energy of the
solitonic shell. This is somewhat speculative but the structure of
diagram \ref{magic_ellipses} gives some support to the conjecture.
Since the circle represents the the limit of the timelike shell
solutions in which the speed of the shell approaches the speed of
light, it is natural that this energy should go to infinity there.

Alternatively we may say the following. In the usual sense, there is
no matter in the vacuum wormhole- the stress-energy tensor is zero
everywhere. There is though `mass without mass'. There are two
disconnected asymptotic regions in the spacetime and no universal
notion of mass. $\tilde Q^a_b$ measures this disagreement between
asymptotic observers. For the thin shell wormhole solutions we have
found, the conserved quantity $\tilde q$ is nicely expressed in
terms of the speed of the shell and $m_+$, $m_-$.

More generally, consider an arbitrary spacetime containing a thin
shell. A geometrical construction of such a spacetime is as follows.
Take two spacetimes which contain submanifolds of codimension 1
which are diffeomorphic to $\Sigma$. We can cut and paste in various
ways. Let us say that we cut out the region to the right of $\Sigma$
in the first manifold and cut out the region left of $\Sigma$ in the
second and make the pasting in that way. Now alteratively we could
cut out the region left of $\Sigma$ on both manifolds, flip the
orientation of one of them and paste in that way. What effect does
this have on the equations of motion of the shell? It is easy to see
that the only effect on the equations of motion for the shell is to
swap the orientation of one of the normal vectors which implies
$\Delta K^a_b \leftrightarrow \tilk^a_b$. Under this transformation,
the five equations (\ref{GB_Junction}-\ref{Very_Long_eqn}) are
unchanged, except that $\Delta Q^a_b$ and $\tilq^a_b$ swap roles. So
$\tilq^a_b$ is what the stress-tensor on the shell would have been
were the orientation of the opposite type, i.e. it measures the
energy difference between configurations related by $\Delta K^a_b
\leftrightarrow \tilk^a_b$. A more detailed study is needed to make
this notion more precise.

\section{Comments}
\label{Comments} Spherically symmetric solitonic thin shell
solutions have been classified in the pure Gauss-Bonnet theory in
five dimensions. The results are summarised in figure
\ref{magic_ellipses}.

We wish to emphasise two points. Firstly, that the pure Gauss-Bonnet
theory, with only the quadratic Lovelock term, is not a physical
theory. Second, that the essential principle that vacuum wormholes
can mimic the effect of a mass, captured here in a simple way,
should indeed generalise to more realistic models.

Let us first expand upon the first point. The pure Gauss-Bonnet
theory has no Newtonian limit: indeed one can see from the form of
the spherically symmetric metric (\ref{spherically_symmetric}) that
$g_{tt} =-1$. A test particle without angular momentum will feel no
central potential.

The theory is also extremely degenerate. One well-known degeneracy
of this theory is the absence of a perturbation theory about
Minkowski space background. Any perturbation about $\Omega^{AB}=0$
is a solution of $\Omega^{AB}\wedge \Omega^{CD}\epsilon_{ABCDE} =0$
to first order. Thus, the second variation of the action about a
Minkowski background is trivial, i.e. the propagator vanishes.

The point particle metric (\ref{spherically_symmetric}) has another
kind of strange degeneracy: it is a solution of the field equations
if $d\Omega^2$ is the metric of a \emph{quite arbitrary}
three-manifold. The three sphere can be replaced with a spheroid, a
hypercube or anything with the topology of a sphere and still be a
solution of a point source with a given mass. This is highly counter
intuitive- one would expect that a single point source would
determine a spherically symmetric spacetime. This very interesting
arbitrariness is something which merits further investigation.

\begin{figure}[]
  \begin{center}
    {\includegraphics[width=.45\textwidth]{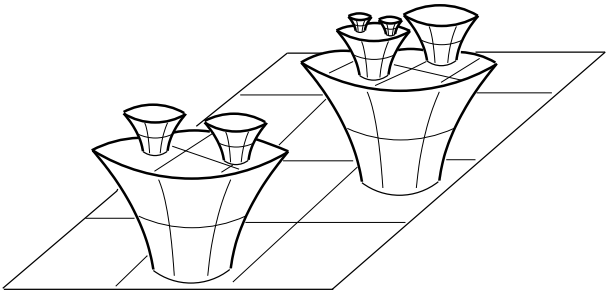}}
    \caption{\small{An exotic vacuum(!) solution of static solitonic shells
    in Minkowski space, showing the extreme degeneracy of Minkowski space
    in the pure Gauss-Bonnet theory.}}
    \label{Trumpets}
  \end{center}
\end{figure}

The solitonic shell solutions are a third example of degeneracy: the
radius at which the static solitons are located is arbitrary. It is
thus possible to have a spacetime composed of different regions with
different $\beta$'s in concentric layers. A single mass can produce
any one of an infinite variety of spacetimes, with the layers being
matched at arbitrary constant radius.
\begin{figure}[]
  \begin{center}
    {\includegraphics[width=.4\textwidth]{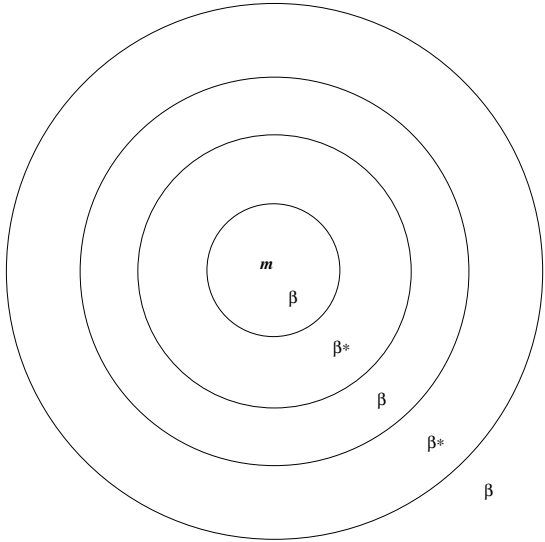}}
    \caption{\small{The hypersurfaces
    separating regions with $\beta$ and $\beta^*$ carry zero energy
    tensor. For static shells these
    betas satisfy
    $\beta^2+\beta\beta^*+(\beta^*)^2=3$.
    A spherically symmetric spacetime around a mass $m$
    is infinitely degenerate. This is true for all masses below
    the critical value $m_{\text{crit}}=2$.
    For more massive particles gravity is simpler!}}
    \label{concentric}
  \end{center}
\end{figure}
The degeneracy is particularly striking for Minkowski space. There
exists a static solution of type II which matches Minkowski space
with a spacetime with mass=2. Now Minkowski space is spherically
symmetric about every point so one can put such solitonic shells
centred about any point. By combining matchings of type IIa) and
IIb) a very exotic vacuum solution can be constructed as in figure
\ref{Trumpets}.

Let us now turn to the second point. The existence of the solitonic
shell wormholes does not depend upon the choice of pure Gauss-Bonnet
theory. The essential feature is the non-linearity of the junction
conditions in the jump in the curvature. Thus, such solutions should
exist in the Einstein-Gauss-Bonnet theory and the feature of mass
without mass should also be exhibited in that theory.  There is a
covariantly conserved quantity which plays the role of $\tilq^a_b$,
as can be seen from equation (\ref{EGB_third}). A thin shell vacuum
wormhole whose throat is a small sphere can be interpreted as a
particle. For such an interpretation to be meaningful, the location
of the throat should be stable. In the pure Gauss-Bonnet theory the
stability analysis leads to a strange situation: the junction
condition gives $\dot{r} =$ constant. The static solutions are
absolutely fixed in place but at an arbitrary radius. This is
special to pure Gauss-Bonnet in which the shell does not accelerate.
In Einstein-Gauss-Bonnet the radius of the shell will be like a
particle in some non-trivial potential. Indeed, such solutions in
the full Einstein-Gauss-Bonnet theory have been found and will be
reported in a separate paper of S.W. with C. Garraffo and G.
Giribet~\cite{Garraffo}.

The solutions for the toy model (pure Gauss-Bonnet) have a simple
structure captured in Fig. \ref{magic_ellipses}. This is especially
simple because of the close relation between the field equation and
the Gauss-Bonnet theorem. Indeed, for the static case the mass in a
given region is exactly equal to the integral of the Euler density
in that region. Let us see how this comes about: For the static
solutions the space-time a trivial product of a spatial
four-manifold with the time direction $M=M^4 \times \mathbb{R}^1$.
The Gauss-Bonnet theorem for the spatial section $M^4$ takes the
form:
\begin{gather*}
 \int_{M^4}\Omega^{AB}\wedge \Omega^{CD}\epsilon_{ABCD0} -
 \int_{\partial M^4} Q_0 =
 -32\pi^2 \chi(M^4)\, .
\end{gather*}
The first term on the left is
 %, by the field equation
 %(\ref{GB_field_equation})
proportional to the integral of $T^0_0$. The boundary term is the
same $Q_0$ which appears in the junction conditions.
%Rearranging this a little
%and setting $16\pi^2 c_2 =1$ we get a formula for the mass in a
%region $M^4$:
%\begin{gather*}
% \text{mass}= \int_{M^4} T_0 =  \chi(M^4) -
% \frac{1}{16\pi^2}\int_{\partial M^4} Q_0\, .
%\end{gather*}
This explains the mass formula (\ref{mass}) and its simple relation
to the junction conditions for the static shell. Also we see why the
solutions of type IIa), the closed universe, have a sum of masses
equal to 2, the Euler number of the spatial manifold, just as for
G.R. in 2+1 dimensions.
 %Note that the boundary term is the same
 %$Q_0$ which appears in the junction conditions. This explains why
 %%the condition for type II solutions takes the form $m_+ + m_- = 2$,
 %twice the Euler number of a disk.}

In the full Einstein-Gauss-Bonnet theory, solitonic shell solutions
should have a much more complicated and rich structure. A useful
intermediate step between our toy model and the full theory is the
case of pure Gauss-Bonnet with cosmological constant. There one can
explore non-trivial features (horizons etc.) in a simple setting.
This is an open problem.
\\

Finally, some comments on the meaning of the solutions are in order.

Spacelike solitonic shells mean lack of determinism, which is rather
a generic feature of Lovelock gravity. For our spherically symmetric
ansatz, these solutions are eliminated for $\beta
> \sqrt{2}$, i.e. $m> 1$. The red circle in Fig. \ref{magic_ellipses}
provides a nice separation between the timelike solitonic shells,
whose behaviour is determined by initial conditions, and the
spacelike (instanton) shells.

 The solutions show that when the
Gauss-Bonnet term is included, wormhole solutions can exist without
an exotic stress tensor as source. Indeed here the stress tensor
vanishes! This is in marked contrast to the situation in Einstein's
theory. Thin shell wormholes were first studied in Einstein gravity
in~\cite{Visser:1989kg}. Also some effects of the Gauss-Bonnet term
as a correction were studied in~\cite{Thibeault:2005ha}. The fact
that wormholes require `exotic matter' in Einstein gravity was
already discussed in~\cite{Morris:1988tu}. Wormhole solutions with
matter source in Einstein-Gauss-Bonnet have been considered in the
past~\cite{Ghoroku:1992tz,Bhawal}.
 There is even another
example of a vacuum wormhole which is already
known~\cite{Dotti:2006cp}. This is a smooth wormhole and exists in
the Lovelock theory with a special choice of coefficients such that
the uniqueness theorem for the Boulware-Deser solution does not
hold~\cite{Charmousis-02,Zegers}. The wormholes found in this
present work are non-smooth, the curvature which defines the throat
is localised in a delta function at the shell.

 The wormhole solutions found here exemplify the concept of
`mass without mass'. It would be interesting to see if, when one
considers the Gauss-Bonnet-Maxwell theory, wormholes can be found
with `charge without charge'. The field equations of the
Gauss-Bonnet theory allow for non-vanishing torsion. Perhaps, by
considering wormholes with torsion, one can create the illusion of a
source for the torsion, `spin without spin'.
\\

Whilst this work was in the final stages, a paper appeared treating
`matter without matter' in Gauss-Bonnet theory~\cite{Maeda-06},
although in a somewhat different context of compactified models in
six and higher dimensions.

\acknowledgments We thank E. Ayon-Beato, G. Dotti, C. Garraffo, A.
Giacomini, G. Giribet, J. Oliva, M. Ramirez, R. Troncoso and J.
Zanelli for useful discussions. S.W. was supported in part by
FONDECYT grant $N^{o}$ 3060016. Institutional support to the
Centro de Estudios Cient\'{\i}ficos (CECS) from Empresas CMPC is
gratefully acknowledged. CECS is funded in part by grants from the
Millennium Science Initiative, Fundaci\'{o}n Andes, the Tinker
Foundation.

\appendix

\section{Some definitions}\label{Defs}
In studying Lovelock gravity, it is useful to introduce the
differential form notation~\cite{Forms_Textbooks}. We introduce the
vielbein $E^A$ and the spin connection $\omega^{AB}$. The curvature
two form is
\begin{gather*}
\Omega^{AB} \equiv d\omega^{AB} + \omega^{A}_{\ C}\wedge \omega^{CB}
= \frac{1}{2} {\cal R}^{AB}_{\ \ CD} \, E^C \wedge E^D.
\end{gather*}
In this notation, the Gauss-Bonnet term is:
\begin{gather*}
{\cal L}_\text{GB} = \Omega^{AB}\wedge \Omega^{CD}\wedge E^F
\epsilon_{ABCDF }\ .
\end{gather*}

In this article, it is assumed that there is no torsion. The spin
connection is the Levi-Civita connection, i.e. an implicit function
of the vielbein. The explicit variation with respect to the spin
connection is a total derivative which contributes nothing to the
field equations. Euler-Lagrange Variation of the action w.r.t. the
vielbein gives the field equation:
\begin{gather}
\Omega^{AB}\wedge\Omega^{CD} \epsilon_{ABCDF } = -2T_F.
\end{gather}
On the right hand side, $T_F$ is the stress-energy 4-form coming
from the matter part of the action. This is the dual of the
stress-energy tensor, which we take to be of the form:
\begin{gather}
 T_{AB} = \left(%
\begin{array}{cc}
  S_{ab} \delta(\Sigma) & T_{Nb} \\
  T_{aN} & T_{NN} \\
\end{array}%
\right)\, .
\end{gather}
We consider a single hypersurface, $\Sigma$ which divides the bulk
space-time into two regions $M_1$ and $M_2$. It is helpful to use
the basis $e_A = (e_a , n)$ adapted to the hypersurface such that
$e_a$ are tangential vectors and $n$ is a normal vector. The
vielbeins $E^A = (E^a , E^N)$ are the dual basis of one-forms.
 We shall assume that
the normal vector $n$ can be spacelike ($\sigma \equiv n\cdot n =
-1$) or timelike ($\sigma = 1$) but not null.

At the hypersurface there is a Levi-Civita connection associated
with each region: $\omega_+^{AB}$ and $\omega_-^{AB}$ respectively.
Let $i^*$ denote the pull-back of differential forms onto $\Sigma$.
Then the intrinsic connection on $\Sigma$ is
\begin{gather*}
\omega_\parallel^{ab}  =i^*\omega^{ab}_+ = i^*\omega_-^{ab}.
\end{gather*}
%This coincides with the Levi-Civita connection induced by the
%intrinsic metric.
Let us define
\begin{gather*}
 \theta^{AB}_{+} \equiv \omega^{AB}_+-\omega^{AB}_{\|}\, .
\end{gather*}
 The second fundamental form on  $\Sigma$ induced by
$M_+$ is $i^*\theta^{AB}_+$ and has components $i^*\theta^{Na}_+ =
i^*\omega^{Na}_+$ and $i^*\theta^{ab}_+ = 0$. It is related to the
extrinsic curvature tensor by:
\begin{gather}\label{Gauss-Weingarten}
 i^*\theta_{\!N}^{\ \, a} = K^a_{\ b} E^b\, .
\end{gather}
Similarly, the second fundamental form induced by $M_-$ is denoted
by $i^*\theta^{AB}_-$.

\section{Proof of the five equations}
\label{GB}

In analysing the field equations, it is useful to introduce a test
field, $\lambda^A$, which is an arbitrary vector valued $1$-form.
The field equations are:
\begin{gather}
\epsilon(\Omega \Omega \lambda) = -2T_A\lambda^A\ ,
\end{gather}
where, to simplify notation, it is convenient to omit indices which
are all contracted with the epsilon tensor. eg:
\begin{gather*}
\epsilon(\Omega \Omega \lambda) : = \Omega^{AB}\wedge
\Omega^{CD}\wedge \lambda^F \epsilon_{ABCDF }\  .
\end{gather*}

The derivation of the five equations by decomposing the field
equations and using the Bianchi identities is purely a technical
one. Applying the five-dimensional Bianchi identities when the
curvature has been decomposed into intrinsic and extrinsic curvature
can be a mess. One elegant way around the problem is to borrow from
the textbook~\cite{Choquet-Bruhat-82} proof of the Chern-Weil
theorem. This has the advantage that the proof generalises easily to
Lovelock theory in arbitrary dimensions. Let $\bo$ be a connection
which interpolates between $\omega_+$ and $\omega_\parallel$.
\begin{gather}
\bo = (1-t)\, \omega_\parallel + t\, \omega_+\ .
\end{gather}
Similarly, one can also interpolate between $\omega_-$ and
$\omega_\parallel$. Then, using the Bianchi identity $D(\omega_t)
\Omega_t = 0$, the following identity can be derived:
\begin{gather}\label{CW}
\epsilon(\Omega_+\Omega_+\lambda) - \epsilon(\Omega_\parallel
\Omega_\parallel \lambda) = 2\int_0^1 dt\ \epsilon\big( D(\bo)
\{\theta_+ \bO\} \lambda\big)
\end{gather}
where $D(\bo)$ is the covariant exterior derivative and $\bO^{AB} =
d\bo^{AB} + \bo^{A}_{\ C}\ \bo^{CB}$ the curvature with respect to
the interpolating connection. The above expression can also be
rewritten in terms of the covariant derivative with respect to the
intrinsic connection:
\begin{gather}
\epsilon(\Omega_+\Omega_+\lambda) - \epsilon(\Omega_\parallel
\Omega_\parallel \lambda) = 2\int_0^1 dt\ \epsilon\big(
D(\omega_\parallel) \{\theta_+ \bO\} \lambda\big) \nonumber\\ +
2\int_0^1 dt\ t \ \theta^{AB}_+\bO^{CD} \epsilon_{ABCDE }\,
\theta^E_{+ F} \lambda^F. \label{start}
\end{gather}
Note that in the above, $D(\omega_\|)$ is a five dimensional
derivative operator. Its projection along the basis of tangential
one-forms is the intrinsic covariant derivative. Its projection
along $E^N$ is just the partial derivative in the normal direction
$n^\mu \partial_\mu$.
\\

%\subsection{The five equations of motion}
We can break down (\ref{start}) into various components:
\\
i) $\lambda$ is a normal 1-form with a tangential vector index, i.e.
the normal-tangent component of the field equations. In this case
the second term on the left and the second term on the right do not
contribute and we obtain
\begin{gather}\label{normal_tangent}
i^*\epsilon(\Omega_+\Omega_+)_a =  i^* D (\omega_\parallel) Q_a^+.
\end{gather}
We have defined the useful quantity $Q_e$:
\begin{gather}\label{def_Q}
Q_e \equiv i^*\, 4\theta_{\!N}^{\ b} \Big( \sigma
\Omega_{\parallel}^{cd} -\frac{1}{3}\theta_{\!N}^{\
c}\theta_{\!N}^{\ d} \Big)\epsilon_{Nbcde}.
\end{gather}
Note that this quantity is closely  related to the boundary term for
the Gauss-Bonnet action for a manifold with boundary
\cite{Myers-87}.
\\
ii) $\lambda$ is a normal 1-form with a normal vector index, i.e.
the normal-normal component of the field equations. In this case we
get
\begin{gather}\label{normal_normal}
i^* \epsilon(\Omega_+\Omega_+)_N = \# {\mathcal H}_\perp ,
\\\nonumber
\# {\mathcal H}_\perp \equiv  i^*\left(-\sigma \Omega_{\|}^{bc}+
\theta_{\!N}^{\ b}\theta_{\!N}^{\ c}\right)\left(- \sigma
\Omega_\parallel^{de} +\theta_{\!N}^{\ d}\theta_{\!N}^{\ e}
   \right)\epsilon_{Nbcde}\, .
\end{gather}
The above formula can be obtained immediately, without reference
to (\ref{start}), by using the Gauss equation.
\\
iii) $\lambda$ is a tangential 1-form with a normal vector index.
This gives the same as case (i) (in the absence of torsion, the
stress tensor is symmetric).
\\
iv) $\lambda$ is a tangential 1-form with tangential vector index.
Integrating this across a region of infinitesimal thickness across
$\Sigma$ gives the known junction
conditions\cite{Davis-02,Gravanis-02}:
\begin{gather}\label{Davis}
\Delta Q_a: = -2 S_a\ .
\end{gather}

We can now substitute the expressions (\ref{normal_tangent}) and
(\ref{normal_normal}) into the field equations. It is most
instructive to evaluate the field equations on the left and the
right and then to consider the sum and the difference:
\begin{align}
i^* D (\omega_\parallel) S_a & = i^*\Delta T_a\ ,
\\
i^* D (\omega_\parallel) \tilq_a & = -2i^*\widetilde{T}_a\ ,
\\
\Delta \#\mathcal{H}_\perp & = -2i^* \Delta T_N\ ,
\\
\widetilde{\#\mathcal{H}}_\perp & = -2i^* \widetilde{T}_N
.\label{last_junction}
\end{align}
The five equations (\ref{Davis}) to (\ref{last_junction}) are the
equations of motion of $\Sigma$ in the absence of torsion, written
in terms of differential forms.
\\

Using equation (\ref{Gauss-Weingarten}) one obtains:
\begin{gather*}
 Q_b = K^{c}_{\ f} \Big(2\sigma R^{de}_{gh} -
\frac{4}{3} K^{d}_{\ g} K^{e}_{\ h} \Big)\epsilon_{Nbcde} E^f \wedge
E^g \wedge E^h\, .
\end{gather*}
Dualising with respect to $\epsilon^{Nafgh}$ one obtains $Q^a_{\ b}$
given by expression (\ref{def_Q_tensor}). Dualising
$\#\mathcal{H}_\perp$ one obtains the scalar:
\begin{align*}
 \mathcal{H}_\perp = \left(\frac{\sigma}{2}R^{ab}_{\ \ ef}
 - K^a_{\ e} K^{b}_{\ f}\right) \left( \frac{\sigma}{2} R^{cd}_{\ \ gh}
 - K^{c}_{\ g} K^{d}_{\ h} \right) \delta^{efgh}_{abcd}
\\
 = \left(\frac{1}{4}
R^{ab}_{ef} R^{cd}_{gh}
  + \frac{\sigma}{2} K^a_{\ e} K^{b}_{\ f} R^{cd}_{\ \ gh}
  \right)
 \delta^{efgh}_{abcd} -\frac{3}{4}K^a_{\ b} Q^{b}_{\ a}\, .
\end{align*}
Thus, dualising the equations (\ref{Davis}) to (\ref{last_junction})
gives the five equations in tensor form. We can combine with the
Einstein term to give the conditions for the general
Einstein-Gauss-Bonnet theory described by the action
(\ref{EGB_action}).
\begin{align}\label{EGB_first}
-c_1 \sigma(\Delta K^a_b - \delta ^a_b\Delta K)
 -\frac{c_2}{2} \Delta Q^a_b & = S^a_b\ ,
\\
 -c_1 \sigma(\Delta K^a_b - \delta ^a_b\Delta K)_{;a}
 -\frac{c_2}{2}\Delta Q^a_{\ b; a} & = \Delta{T}^N_{\ \, b}\, ,
\\
 -c_1\sigma(\tilk^a_b - \delta ^a_b\tilk)_{;a}
  -\frac{c_2}{2} \tilq^a_{\ b; a} & = \widetilde{T}^N_{\ \,b}\,
  ,\label{EGB_third}
\\
 \frac{c_1 \sigma}{4}
  \Delta K^a_c \tilk^b_d \,
  \delta^{cd}_{ab} -\frac{c_2}{2}\Delta\mathcal{H}_\perp
  & = \Delta T^N_N\ ,
\end{align}
\begin{multline}
-c_0 -c_1\left( R - \frac{\sigma}{4}\left\{
  \tilk^a_c \tilk^b_d + \Delta K^a_c
  \Delta K^b_d
\right\}\delta_{cd}^{ab}\right)
  \\
-\frac{c_2}{2}\widetilde{\mathcal{H}_\perp}= \widetilde{T}^N_{\
\,N} \, .\label{EGB_fifth}
\end{multline}

%Finally, note that the method of decomposing the field equations
%using equations (\ref{CW}) and (\ref{start}) generalises
%straightforwardly to higher order Lovelock theory. Let us consider
%the $n$th order Lovelock term in dimension greater than $2n$. One
%finds that
%\begin{gather}
%i^*\left(\epsilon(\Omega_+\cdots \Omega_+ E \cdots E)_a
%-\epsilon(\Omega_\parallel \cdots \Omega_\parallel E \cdots
%E)_a\right)\nonumber \\ = i^*\left(n \int_0^1 dt\ \epsilon\big(
%D(\omega_\parallel) \{\theta_+ \bO\cdots \bO\} \big)_a
%\right)+(\cdots)\,
%\end{gather}
%where the term $(\cdots)$ contains no normal derivatives of the
%discontinuous components of the curvature and so contributes nothing
%to the junction conditions.

We show now that the junction condition (\ref{Davis}) is well
defined. The thin-shell limit is well defined in the following
sense: starting from a thick shell, in the limit that its thickness
becomes zero, the results are insensitive to the way in which the
limit is taken. This has been discussed at some length in the
literature (e.g. \cite{Deruelle:2003ur}) but it is worth giving a
precise statement of this here, since the subject still causes some
confusion. To see this more explicitly, let us define a family of
metrics $g_{AB}^{(\alpha)}$, parameterised by a positive number
$\alpha$, which describe a thick shell of characteristic thickness
$\propto 1/\alpha$. We can foliate the neighbourhood of the shell
into tangential slices and a normal vector $N^{(\alpha)}$. Let us
define $\theta^{(\alpha)} \equiv \omega^{(\alpha)} -
\omega_\parallel$, where $\omega_\parallel$ is the intrinsic
connection induced on the slice. In the limit $\alpha \to \infty$,
the 1-form $\theta^{(\alpha)}$ tends towards something
discontinuous: it is equal to $\theta_+$ in one region and
$\theta_-$ in the other region. So the components of $\theta$ become
discontinuous.

Let us now look at what happens to the field equations,
$\epsilon(\Omega\Omega\lambda)=-2T_A \lambda^A$. From the identity
(\ref{CW}) we obtain
\begin{gather}
\epsilon(\Omega\Omega\lambda) - \epsilon(\Omega_\parallel
\Omega_\parallel \lambda) = d \Big\{2\int_0^1 dt\ \epsilon\big(
\theta \bO \lambda\big) \Big\} \nonumber\\ + 2\int_0^1 dt\
\theta^{AB}\bO^{CD} \epsilon_{ABCDE }\, D(\bo) \lambda^E.
\label{start2}
\end{gather}
 We
integrate the identity (\ref{start2}) over the thick shell. The
potentially singular terms are those which contain the normal
derivative of $\theta^{(\alpha)}$. Everything else is smooth or
remains finite. The first term on the r.h.s. of (\ref{start2}) gives
the junction condition (\ref{Davis}). The claim is that this term
contains all the singular terms. From the second term on the r.h.s.
of (\ref{start2}) the normal derivative of $\theta^{(\alpha)}$
appears as
\begin{equation}\label{dangerous}
\epsilon_{ABCDE}\, \theta_a^{~AB} \, \partial_N \theta^{~CD}_b \:E^N
\!\!\wedge\! E^a \!\wedge\! E^b \!\wedge\! D(\bo)\lambda^E,
\end{equation}
using an adapted frame $(E^N,E^a)$ on the foliation. The index
$\alpha$ will be dropped from now on. Now $\theta_a^{~AB}$ are the
components of the second fundamental form of a slice in the
foliation: one of the indices $A,B$ is a normal index. Thus two
indices contracted to the antisymmetric symbol $\epsilon_{ABCDE}$
are normal. The quantity (\ref{dangerous}) vanishes identically.
The integral of the second term in the r.h.s. of (\ref{start})
goes to zero for $\alpha\to\infty$. The discontinuity of $\theta$
is contained in a total derivative. The singular part of the field
equations is well defined as a Dirac $\delta$ distribution.

Let's say a few more words. The equations of motion
$\epsilon(\Omega\Omega\lambda)$ have a singular term of the form
$\psi \phi \partial_N \chi$, where $\psi$ $\phi$ and $\chi$ are
three different components of $\theta$. In general, these components
will converge in the weak sense to a Heaviside type of distribution
$H$ in different ways, so the product does not in general tend
towards an unambiguous distribution (i.e. it is not exactly of the
form $H^2(x) \delta(x) = \delta(x)/3$, where $\delta$ is the Dirac
$\delta$ distribution, as would happen with a single function
discontinuous at $x=0$). However, as discussed above, the simple
fact is that they always appear in a combination which is a total
derivative $\partial_N(\psi \phi \chi)$. This product of functions
in the brackets is a single function which converges to $H$. Its
derivative is unambiguously defined as a distribution in the limit
as $\delta$. A corollary of this and of the comments above is that
the integral of the field equations of a physically thin shell is
well described by these equations, i.e. if we allow for shell to
have a little thickness then in a first order approximation the
results do not change if we change the configuration in the
interior. The integrated stress tensor $S_{ab}$ is unambiguously
related to $\theta_+$ and $\theta_-$, the values of $\theta$ on each
side of the shell.

\end{document}